\documentclass[12pt,a4paper,fleqn]{article}
\usepackage[utf8x]{inputenc}
\usepackage[english]{babel}
\usepackage{slashed}
\usepackage[T1]{fontenc}
\usepackage{cite}
\usepackage{float}
\usepackage[font=small,labelfont=md]{caption}
\usepackage{hyperref}
\usepackage{amsmath,amsthm,amssymb,latexsym}
 \usepackage[english]{babel}
\usepackage[dvips]{graphicx}
\usepackage{amsfonts}
\usepackage{textcomp}
\usepackage{subfig}
\usepackage{pstricks}
\usepackage{authblk}
\usepackage[normalem]{ulem}
\usepackage{color}
\usepackage[title,titletoc,toc]{appendix}
\usepackage{multirow}

\def\be{\begin{equation}}
\def\ee{\end{equation}}
\def\bea{\begin{eqnarray}}
\def\eea{\end{eqnarray}}
\def\bear{\begin{array}}
\def\eear{\end{array}}

\title{The right  top coupling in the aligned two-Higgs-doublet model }

\author[1,2]{Cesar Ayala\thanks{cesar.ayala@uv.es}}
\author[3]{Gabriel A. Gonz\'alez-Sprinberg\thanks{gabrielg@fisica.edu.uy}}
\author[4]{R.  Martinez\thanks{remartinezm@unal.edu.co}}
\author[1]{Jordi Vidal\thanks{vidal@uv.es}}

\affil[1]{{\small\it  Departament de F\'\i sica Te\`orica, Universitat de Val\`encia {\rm \&} Instituto de F\'\i sica Corpuscular (IFIC), Centro Mixto Universitat de Val\`encia-CSIC, E-46100 Burjassot, Val\`encia, Spain.}}
\affil[2]{{\small\it  Department of Physics, Universidad T\'ecnica Federico Santa Mar\'ia, Casilla 110-V, Valpara\'iso, Chile.}}
\affil[3]{{\small\it  Instituto de F\'\i sica, Facultad de Ciencias, Universidad de la Rep\'ublica, Igu\'a 4225, Montevideo 11600, Uruguay.}}
\affil[4]{{\small\it  Departamento de F{\'\i}sica, Universidad Nacional de Colombia, Bogot\'a Distrito Capital, Colombia}}

\begin{document}

\maketitle
\vspace {-13cm}
\noindent IFIC/16-76 \hfill  FTUV-16-1123.8038
\vspace {12cm}

\begin{abstract}
We compute the right top quark coupling in the aligned two-Higgs-doublet model. In the Standard Model the real part of this coupling is dominated by QCD-gluon-exchange diagram, but the imaginary part, instead, is purely electroweak at one loop.  Within this model we show that values for the imaginary part of the coupling up to one order of magnitude larger than  the electroweak prediction can be obtained. For the real part of the electroweak contribution we find that  it can be up to three orders of magnitude larger than  the standard model one.  We  also present detailed results of the one loop analytical computation.

\end{abstract}

\section{Introduction}

In 2015 the LHC center-of-mass energy has reached 13 TeV. By the end of 2016 the LHC will be close to a peak luminosity of $1.5\times 10^{34} \, cm^{-1} \, s^{-1}$, with an  integrated luminosity of $38 fb^{-1}$. After 2020, several components of the accelerator will reach  the  radiation damage or reliability limit so that,  by 2024 the LHC will have to be upgraded to the High-Luminosity LHC (HL--LHC), which is expected to accumulate over the next 10 years an impressive integrated luminosity of $3000\ f b^{-1}$ at energies close to 13-14 TeV \cite{Selvaggi:2015sdf,Barletta:2013ooa}. The CMS and Atlas experiments have already collected millions of top quark pairs and single top events but in this scenario of very high luminosity, they will detect billions of them in the future. Besides,  next generation of colliders, such as CLIC,  will eventually be built and it is expected that the top quarks physics will enter in an era of high precision.  The top quark  is the only quark that decays weakly before hadronization and, up to now, only one decay mode, $t \rightarrow b W^+$, is known. It was detected for the first time at TEVATRON\cite{Abe:1995hr,Abachi:1995iq} where  many of its  physical properties were first measured and also some limits on the anomalous $tbW$ couplings were set \cite{Deterre:2012vn, Abazov:2011pm,Abazov:2008sz}. 

Top quark physics is considered as one of the gateways to new physics \cite{Bernreuther:2008ju,Bardhan:2016txk,Bernreuther:2015yna} and the study of its decay properties at the LHC \cite{Schilling:2012dx,Bernardo:2014vha, Hawkings:2015ega,Cristinziani:2016vif} is being extensively investigated by the ATLAS and CMS collaborations \cite{Calvet:2014fxa,Pagano:2013goa}. The determination of other couplings of the top quark, such as the  chromoelectric and chromomagnetic of the $ttg$ (top-top-gluon) vertex  has been recently suggested \cite{Gaitan:2015aia} as a window for new physics, in  the two-Higgs-doublet model ({\it 2HDM}) framework with a $CP$-violating potential. The study of the different helicity components of the $W$ in the top decay has been also proposed to investigate the $tbW$ Lorentz vertex structure \cite{delAguila:2002nf}. In recent works \cite{AguilarSaavedra:2010nx,Drobnak:2010ej,Rindani:2011pk,Prasath:2014mfa,Cao:2015doa,Hioki2016128} it has been shown that a precise  determination of the  Lorentz form factors of the $tbW$ vertex  can be done with a suitable choice of observables built from longitudinal and transverse helicities of the $W$ coming from the top decay. 

The enormous amount of collected data by the LHC (and in the future by the HL--LHC) will determine the  complete structure of the $tWb$ vertex, with a precise determination of the properties of top quark couplings to the $W$ boson and to the $b$ quark.

The most general parametrization of the on-shell vertex needs four couplings. In the Standard Model (SM) the left coupling $V_L$ is not zero and takes a value close to one \cite{Olive:2016xmw}. The other three are zero at tree level: the chiral $V_R$ coupling, and the left $g_L$ and right $g_R$ anomalous tensorial couplings.  This is not the case in extended models where, in addition, some of these couplings can also be sensitive to new CP-violation mechanisms. The measurement of the  two tensorial couplings $g_{L,R}$ at the  LHC was investigated  in ref. \cite{MorenoLlacer:2014tca}. The values of $g_{L,R}$ within the SM,   the {\it 2HDM} and other extended models  where recently calculated in refs. \cite{GonzalezSprinberg:2011kx,Duarte:2013zfa,Bernreuther:2008us} and they will not be considered in this paper. The right top coupling $V_R$ was computed in the SM  at leading order in ref. \cite{Gonzalez-Sprinberg:2015dea}.

The LHC observables considered  in the literature are not, in general,  very sensitive to the right coupling $V_R$. This is due to the fact that in the lagrangian  the $V_R$ coupling has the same parity and chirality properties than the  leading coupling $V_L$, so that the observables receive contributions from both terms. Some of these observables are the angular asymmetries in the $W$ rest
frame\cite{Lampe:1995xb,delAguila:2002nf,AguilarSaavedra:2006fy, AguilarSaavedra:2010nx}, angular asymmetries in the top  rest frame \cite{AguilarSaavedra:2006fy,Grzadkowski:1999iq,Godbole:2006tq,AguilarSaavedra:2010nx} and spin correlations
\cite{Stelzer:1995gc,Mahlon:1995zn,AguilarSaavedra:2006fy,AguilarSaavedra:2010nx}. In ref. \cite{Gonzalez-Sprinberg:2015dea} some of these observables were redefined in order to be directly proportional to the coupling we are interested in, $V_R$, in such a way as to cancel the leading $V_L$ contribution to them. Then, these observables are directly sensitive to  $V_R$ and can be  an important tool in order to search for new physics contributions to this coupling. 

A simple and widely studied extension of the  electroweak theory is to  consider a second scalar
doublet added to the SM.  However, tree level flavour changing neutral currents (FCNC) arise unless new 
hypothesis are introduced.  A  solution to this issue is the
 aligned two Higgs doublet model {\it A2HDM} \cite{Pich:2009sp}, where the two Yukawa matrices coupled to the same type of right-handed fermion are aligned in flavour space. Then, no FCNCs  appear at tree level. Besides, most of the popular 
versions of the  {\it 2HDM} are reproduced with particular choices of the {\it A2HDM} parameters.   In this paper we present a detailed calculation of the new contributions to the $V_R$ top right coupling in the general framework of the {\it A2HDM}.

This work is organized as follows. In the next section we briefly review the {\it A2HDM}, introducing the notation used in the paper and presenting the current limits that constraint the parameters of the model. In section 3 we define the vertex parametrization and show the details of the computation of the different contributions to the right vector coupling $V_R$ within the {\it A2HDM}. In section 4 we investigate the sensitivity of the $V_R$ coupling to the scalar mixing angle and alignments parameters, for a CP-conserving scalar potential. We show  the results obtained for values of the parameters of the model and masses
of the new particles so as to cover the meaningful parameter space of the model. The results for {\it 2HDM} Type-I and II  are also shown. We present our conclusions in section 5.

\section{The aligned two-Higgs-doublet model}\label{A2HDMov}
The {\it 2HDM} extends the SM by adding a second scalar doublet $\phi_2(x)$ with the same hypercharge $Y=1/2$ \cite{Lee:1973iz,Branco:2011iw}. Similarly to what happens in the SM, after symmetry breaking, the neutral components of the two doublets get non zero vacuum expectation values $\left<\phi_i\right>_{i=1,2}^T=(0,\frac{v_i}{\sqrt{2}}e^{i\theta_i})$. 

The so called Higgs basis  $(\Phi_1(x),\Phi_2(x)$  is obtained through a rotation  of the $\phi_1(x)$,  $\phi_2(x)$ states given by the angle $\beta$ (defined as $\tan\beta=\frac{v_2}{v_1}$), in such a way that only one of the doublets ($\Phi_1(x)$) gets a non-zero expectation value $v=\sqrt{v_1^2+v_2^2}$. 

In this basis, the three components of the doublets can be written as 
\begin{equation}
\Phi_1(x)=    \left(\begin{array}{c} G^+(x)\\ \frac{1}{\sqrt{2}}\left(v+S_1(x)+i\, G^0(x)\right)\end{array}\right),\quad 
\Phi_2(x)=    \left(\begin{array}{c} H^+(x)\\ \frac{1}{\sqrt{2}}\left(S_2(x)+i\, S_3(x)\right)\end{array}\right)\label{eq:1}
\end{equation}
where $G^0(x)$ and $G^\pm(x)$ correspond to the three would-be Goldstone bosons of the SM, $H^\pm(x)$ are two new charged scalar fields and $\{S_i(x)\}_{i=1,2,3}$ are three neutral scalars with no defined mass. To get the three mass eigenstates as a linear combination of the later three scalars one has to perform an orthogonal transformation $\mathcal R$ so that the new three mass eigenstates, $\{ \varphi_{i}(x) \}_{i=1,2,3} =\{h(x),H(x),A(x)\}$, can be written as
\begin{equation}
\varphi_{i}(x)= \mathcal{R}_{ij}S_{j}(x) \; ; \,\;\, i, j = 1, 2, 3.
\end{equation}
The particular form of the potential will define the matrix $\mathcal{R}$ and the  structure of the scalar mass matrix and mass eigenstates. If the potential is CP-conserving, the CP-even states $\{S_1(x),S_2(x)\}$ will not mix with the CP-odd one ($S_3(x)$) 
so that:
 \begin{eqnarray}\label{matrix}
H(x)&=&  \cos\gamma\, S_1(x) +\sin\gamma \, S_2(x)\nonumber\\
h(x)&=& -\sin\gamma\, S_1(x)+\cos\gamma\, S_2(x)\nonumber\\
A(x)&=& S_3(x)
\end{eqnarray}
where $\gamma$ is the neutral scalars mixing angle.

The most general Yukawa Lagrangian, with standard fermionic content will have different couplings to $\Phi_1(x)$ and $\Phi_2(x)$ doublets. It means that when one diagonalizes the fermionic mass matrices -in the Higgs basis-  this transformation will no diagonalize the fermion-scalar Yukawa matrices. The Yukawa lagrangian can then be written as
\begin{eqnarray}
\label{eq2:LYukawaA2HDM}
 \mathcal{L}_{Y}&=& -\frac{\sqrt{2}}{v} 
 \Big\{\bar{Q}_L(x)\left[{\cal M}_d\Phi_1(x)+{\cal Y}_d\Phi_2(x)\right]d'_R (x) \nonumber\\
 &&+\ \bar{Q}_L(x)\left[{\cal M}_u
 \tilde{\Phi}_1(x)+{\cal Y}_u\tilde{\Phi}_2(x)\right]u'_R(x)\\
&& +\ \bar{L}_L(x)\left[{\cal M}_l\Phi_1(x)+{\cal Y}_l\Phi_2(x)\right]l'_R (x)+ h.c.\Big\}\nonumber ,
\end{eqnarray}
where $\tilde{\Phi}_i(x)=i\tau_2\Phi_i^*(x)$, all fermionic fields, $Q_L(x)$, $L_L(x)$, $d'_R(x)$, $u'_L(x)$ and $l'_R(x)$, are three-dimensional vectors in the flavour space, ${\cal M}_f$ ($f=d,u,l$) are the non-diagonal $3\times3$ fermion mass matrices, and ${\cal Y}_f$ are the fermion-scalar Yukawa couplings that are, in general, also non-diagonal. The rotation to the fermionic mass eigenstates ($d(x)$, $u(x)$, $l(x)$, $\nu(x)$) which diagonalizes the mass matrices ${\cal M}_f$ will, in general, not diagonalize simultaneously the Yukawa matrices ${\cal Y}_f$, so that they will introduce FCNC at tree level. Among the different approaches to avoid this unwanted effect we choose the one that, before diagonalization,  makes both Yukawa matrices -${\cal M}_f$ and ${\cal Y}_f$, for each type of right handed fermions-  proportional to each other (alignment in the flavour space). Then, they can be simultaneously diagonalized and the diagonal Yukawa matrices satisfy the relations:
\begin{equation}
    Y_j=\varsigma_j \, M_j,\; i=d,l \quad  Y_u=\varsigma_u \, M_u,\quad \varsigma_f^*=\frac{\xi_f-\tan\beta}{1+\xi_f\tan\beta},\quad f=u,d,l
\end{equation}
with $\xi_f$ being an arbitrary complex number and $M_f$ ($f\equiv u,d,l$)  diagonal mass matrices. This is the so called {\it A2HDM}. It has the advantage that for different values of the $\xi_f$ parameter (see \cite{Pich:2009sp}) it reproduces the {\it 2HDM} with discrete $Z_2$ symmetries, Type-I, II, X, Y and inert model. Obviously if the $\varsigma_f$ are taken to be arbitrary complex numbers the Lagrangian incorporate new sources of CP-violation.

The Yukawa lagrangian can be then written as:
\begin{align}\label{eq1:LYukawaA2HDM}
 \mathcal{L}_{Y}= &-\frac{\sqrt{2}}{v} H^{+}(x)\bar u(x)[\varsigma_{d}VM_{d}P_{R}-\varsigma_{u} M_{u}VP_{L}]d(x)\nonumber \\
 \qquad {} &
 -\frac{\sqrt{2}}{v}H^{+}(x)\varsigma_{l}\bar \nu(x) M_{l}P_{R}l(x)\nonumber \\
 \qquad {} &
 -\frac{1}{v}\sum_{i,f}\varphi_{i}(x)y^{\varphi_{i}}_{f}\bar f(x)M_{f}P_R f(x)+ h.c.\ ,
\end{align}
where $V$ is the Cabibbo-Kobayashi-Maskawa matrix and  $P_{R,L}\equiv\frac{1}{2}(1\pm\gamma_{5})$ are the chirality projectors.

The neutral Yukawa terms are flavor-diagonal and the couplings $y^{\varphi_{i}}_{f}$ ($\varphi_i=h,H,A$) are proportional to the corresponding elements of the neutral scalar mixing matrix $\mathcal{R}$:
\begin{eqnarray}
y^{\varphi_{i}}_{d,l}&=& \mathcal{R}_{i1}+(\mathcal{R}_{i2}+i\mathcal{R}_{i3})\, \varsigma_{d,l}\, ,\nonumber \\
y^{\varphi_{i}}_{u}&=& \mathcal{R}_{i1}+(\mathcal{R}_{i2}-i\mathcal{R}_{i3})\, \varsigma^*_{u}
\end{eqnarray}
that, in the particular case of a CP-conserving potential can be written as:

\begin{equation}
\begin{array}{lcl}
 y^{H}_{d,l}&=& \cos\gamma +\sin\gamma \, \varsigma_{d,l}\, ,  \\
   y^{h}_{d,l}&=& -\sin\gamma +\cos\gamma\, \varsigma_{d,l}\, ,  \\ 
   y^{A}_{d,l}&=& i \, \varsigma_{d,l}\, ,
\end{array}
\qquad\quad
\begin{array}{lcl}
  y^{H}_{u}&=& \cos\gamma +\sin\gamma \, \varsigma^*_{u}\, ,  \\
   y^{h}_{u}&=& -\sin\gamma +\cos\gamma\, \varsigma^*_{u}\, ,  \\ 
   y^{A}_{u}&=& -i \, \varsigma^*_{u}\, .
\end{array}
\end{equation}

Then, the CP-conserving {\it A2HDM} contains 10  real parameters: the three complex alignment constants $\varsigma_{u,d,l}$, the three scalar masses $m_{A,H,H^\pm}$, and the scalar mixing angle $\gamma$. 
We will assume that the light CP-even Higgs $h$ is the SM-like Higgs with a mass of $125.09$~GeV \cite{Olive:2016xmw}.  
The other  parameters have not yet been measured  and they can be constrainted  by indirect phenomenological and theoretical arguments. 

The presence of a charged Higgs is a  signature of the model that allows some constraints coming from the phenomenolgy associated. In ref. \cite{Jung:2010ik} combined bounds on $\varsigma_{u,d,l}$ and $m_{H^\pm}$ are obtained from: a) tau decays, $|\varsigma_{l}|/m_{H^{\pm}} \leq 0.40$~GeV$^{-1}$, and b) a global fit to the tree leptonic and semi-leptonic decays of pseudoscalar mesons, $|\varsigma_{u} \varsigma^{*}_{l}|/m^{2}_{H^{\pm}} \lesssim 0.01$~GeV$^{-2}$ and $|\varsigma_{d} \varsigma^{*}_{l}|/m^{2}_{H^{\pm}} \lesssim 0.1$~GeV$^{-2}$.
Bounds can be improved by looking at loop-induced processes, $Z\rightarrow b\bar{b}$, $B^0$--$\bar{B}^0$ and $K^0$--$\bar{K}^0$ mixing, and $\bar{B}\rightarrow X_s\gamma$, assuming that the dominant new-physics corrections to  the observables are those generated by the charged scalar; then $|\varsigma_u|<1.91$ for $m_{H^\pm}=500$~GeV \cite{Jung:2010ik,Jung:2012vu}. 

Bounds on $\varsigma_d$ are more difficult to get from phenomenology so an upper bound as big as $|\varsigma_d|\leq 50$ can be used \cite{Jung:2012vu}. Studies of the radiative decays $\bar{B}\rightarrow X_{s,d}\, \gamma$, show that the combination $|\varsigma_{u}^*\varsigma_{d}|$ is strongly correlated with the mass of the scalar charged boson $m_{H^\pm}$, thus one find that $|\varsigma_{u}^*\varsigma_{d}|\leq 25$ for $m_{H^\pm}\in(100,500)$~GeV \cite{Jung:2012vu,Jung:2010ik}. More constraints  on the $\varsigma_u$--$\varsigma_d$ plane can also be set from $\bar{B}$ decays and are given in ref. \cite{Jung:2012vu,Jung:2010ik}

Recently, direct searches of light  charged scalar Higgs in  $t\rightarrow H^+\, b$ decay in ATLAS and CMS  \cite {Chakraborty:2015qja}  give an upper bound \cite{Celis:2013ixa} on the combination $|\varsigma_{u}^*\varsigma_{d}|$ that excludes part of the allowed regions constrained by $\bar{B}$ decays.

All these limits put  constraints on the parameter space of the  model. In this paper we only consider  the ones that are related to the top physics. 

\section{\texorpdfstring{$V_R$}{Vr} top coupling in the {\it A2HDM}}\label{gesA2HDM}

The most general Lorentz structure of the amplitude $\mathcal{M}_{tbW}$, for on-shell particles, in  the $t(p)\rightarrow b(p') W^{+}(q)$ decay is:
\begin{eqnarray}
\mathcal{M}_{tbW}&=& - \frac{e}{\sin\theta_{w}\sqrt{2}}\epsilon^{\mu*} \, \times \nonumber \\
&& \overline{u}_{b}(p')\left[\gamma_{\mu}(V_{L} P_{L}+V_{R} P_{R})+ 
\frac{i \sigma_{\mu\nu}q^{\nu}}{m_{W}}( g_{L} P_{L}+ g_{R} P_{R})\right]u_{t}(p),\label{gesdef}
\end{eqnarray}
where the outgoing $W^{+}$ momentum, mass and polarization vector are $q=p-p'$, $m_{W}$ and $\epsilon^{\mu}$, respectively.
The couplings are all dimensionless; $V_{L}$ and $V_{R}$ parametrize the left and right vector couplings while $ g_{L}$ and $ g_{R}$ are the so called left and right anomalous tensor couplings, respectively.

In an effective Lagrangian approach these couplings arise as  contributions  of  low energy non-renormalizable lagrangian terms, originated in  a high energy theory. 
This approach assumes that the new physics spectrum is  well above the electroweak (EW) energy scale \cite{Buchmuller:1985jz, AguilarSaavedra:2008zc, Kane:1991bg}.

The couplings $V_R$,  $g_R$ and  $g_L$ are  zero at tree level within the SM, and  $V_L$  is given by the Kobayashi-Maskawa matix element $V_L=V_{tb} \simeq 1$ \cite{Agashe:2014kda}. The values of the anomalous tensor couplings at one loop have been calculated in ref. \cite{GonzalezSprinberg:2011kx} for the SM,
and in ref. \cite{Duarte:2013zfa} for a general {\it A2HDM}. 

The SM contribution to $V_R$ has   been  calculated in ref. \cite{Gonzalez-Sprinberg:2015dea}. There, the QCD one loop gluon exchange and the one loop contribution from  the EW sector of the SM have been explicitly calculated.
For the values of the standard masses and couplings given in  \cite{Agashe:2014kda}, they are:
\begin{equation}
V_R(QCD)=2.68\times 10^{-3}, \qquad V_R(EW)= (-0.015+8.92\,i)\times 10^{-5}.
\end{equation}

Note that, as can be seen from ref. \cite{Gonzalez-Sprinberg:2015dea},   the EW contribution to the real part of the $V_R$ coupling from most of the EW diagrams is of the order of $10^{-5}$ but, due to accidental cancellations among them, the final result is two orders of magnitude smaller. In fact this real part, within the precision of our calculation and considering the uncertainties of the data used, is compatible with zero, at $10^{-7}$ precision.\footnote{Notice that the result quoted here differs (even in sign) from the one of ref.\cite{Gonzalez-Sprinberg:2015dea}. As explained, this is so for two reasons: 1) the set of PDG values used here for the SM parameters is different and, 2) the accidental cancellation among diagrams makes the result very sensitive to these  values, and consequently, the final result is not well determined  and strongly depends on small changes on the SM masses and couplings within the experimental errors  given in \cite{Agashe:2014kda}.} The imaginary part, instead, remains of order $10^{-5}$, and it is purely EW.

\begin{figure}[H]
\begin{center}
\includegraphics[width=8cm]{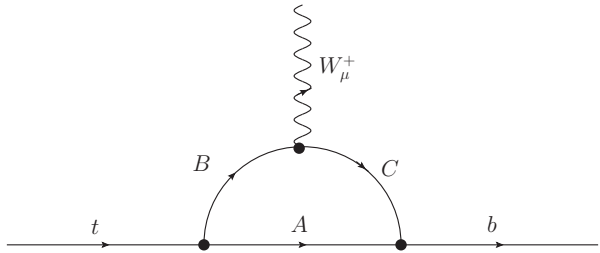}
\end{center}
\caption{One-loop contributions to the $V_R$ coupling in the $t\rightarrow b W^+$ vertex.}
\label{figura1}
\end{figure}

In the {\it 2HDM}, the couplings structure of the $tbW$ remains unchanged at tree level. However,   at one loop, in addition to the usual particle contents of the SM, the three new neutral scalars $h$, $H$ and $A$, and the new charged scalars $H^{\pm}$ of the {\it 2HDM} may circulate in the internal lines of the loop and new contributions to the $V_R$ coupling arise.
The structure of the one loop diagrams contributing to the $V_R$ top right-coupling is given in figure \ref{figura1}.

We denote each diagram by the label $ABC$ according with the particles running in the loop. In table \ref{tabla2} we shown the 17 new diagrams to be considered, ordered by the position (A, B, or C) of the neutral scalars $\varphi_i$, where  $\varphi_i$ stands for one of the neutrals $h$, $H$  and $A$  in the diagram types from (1) to (3), while for diagrams types  (4) to (7), $\varphi_i$ runs only for the neutral scalar bosons $h$ and $H$. It is important to notice that diagrams type (5) and (7) always have an imaginary part while, depending on the mass of the new scalar charged Higgs ($H^\pm$), diagrams type (2) may or may not develop it.

\begin{table}[ht]
\centering
\begin{tabular}{|c|c|l|}
\hline
Type& \multicolumn{2}{|c|}{Particles in the loop $ABC$}\\ \hline
(1)&$t\,  \varphi_i \, H^-$&\multirow{3}{*}{$\varphi_i=h,H,A$}\\ \cline{1-2} 
(2)& $b\, H^+\, \varphi_i$&\\ \cline{1-2}
(3)& $\varphi_i\, t\, b\quad$&\\ \hline
(4)& $t\, \varphi_i\, W^-$&\multirow{4}{*}{$\varphi_i=h,H$}\\ \cline{1-2}
(5)& $b\, W^+ \, \varphi_i $&\\ \cline{1-2}
(6)& $t\, \varphi_i\, G^-$&\\ \cline{1-2}
(7)& $b\, G^+\, \varphi_i$&\\ \hline
\end{tabular}
\caption{\label{tabla2} Classification of the new the Feynman diagrams by the particles circulating in the loop.}
\end{table}

Chirality imposes that all the contributions are proportional to the bottom mass and can be written as:
\begin{equation}
V_R^{ABC} = \alpha\; V_{tb}\; r_b\, I^{ABC},
\label{vr}\end{equation}
where $r_b=m_b/m_t$ and $I^{ABC}$ is the Feynman integral corresponding to the given diagram. In appendix \ref{app:AppendixA} we give the analytical expressions of all these integrals, for the diagrams shown in table \ref{tabla2}.

The $V_R$ coupling depends on the scalar mixing angle $\gamma$ and on the alignment parameters $\varsigma_{u}$ and $\varsigma_{d}$. The mass dependence is parametrized by the dimensionless variable $r_{X}=m_{X}/m_{t}$, where $m_X$ is the mass of the particle $X$ circulating in the loop.   
For the neutral scalar masses above the TeV scale, the Feynman integrals give negligible values when compared to the SM contributions. However, the $V_R$ coupling is very sensitive to the new particles masses when they take lower values.

As  in the SM, some of the diagrams  are  ultraviolet divergent, but we know that the total result must be finite. In appendix \ref{app:AppendixA} it can be seen that the sum of diagrams  (3), (6) and (7),  to the  SM diagrams $G^0tb$, $tG^0G^-$ and $bG^+G^0$, respectively, cancel all the ultraviolet divergences and the total result is finite. This fact has been also used as a test of our analytical calculation\footnote{ The logarithmic terms in the expressions given in appendix \ref{app:AppendixA} are the finite contributions coming from the sum of the divergent  part of each of the diagrams evaluated. }.

We recover the SM expressions from the {\it A2HDM} just by taking the  $\varsigma_{u,d} \rightarrow 0$ limit and setting $\gamma = - \pi/2$, in such a way that the neutral scalar $h$  has the same couplings as the SM Higgs boson.  In that limit we  explicitly checked  that the contributions to the top right-coupling  in the {\it A2HDM} --diagrams type (3) to (7)--   are identical to the corresponding ones in the SM obtained in ref.\cite{Gonzalez-Sprinberg:2015dea}. 

\section{Results}

In this section we present the  one loop corrections to the top right-coupling $V_R$  in the {\it A2HDM}. As already stated, these corrections depend on the alignment parameters $\varsigma_{u,d}$, the scalar mixing angle $\gamma$ and on the masses of the new particles: 2 neutrals scalars $h$ and $H$, one axial $A$, and two charged scalars $H^\pm$. We  write the alignment parameters as:
\begin{equation}
 \varsigma_{u}=\rho_{u} e^{i\theta_{u}}, \qquad \varsigma_{d}=\rho_{d} e^{i\theta_{d}},
\end{equation}
and we investigate  separately the effects of modulus and phases on the top $V_R$ right-coupling. In addition to the masses of the new particles we  have five free parameters: $\rho_{u}$, $\rho_{d}$, $\theta_{u}$, $\theta_{d}$ and the mixing angle $\gamma$.
 
We chose different sets of  values for the masses  of the new neutral and charged scalar particles;  the  scenarios we consider are shown in table \ref{escenariosmasas}. The new scalar masses are taken to be of the order of $10^2$~GeV \cite{Aaltonen:2011rj, Abazov:2011ix}. In the framework of {\it 2HDM} and  under certain assumptions on its dominant decays, the charged scalar mass, $m_{H^{+}}$, is excluded to be below $85$~GeV by LEP data \cite{Abbiendi:2013hk}. Then, it can take values below  the top quark mass,  so that  the decay $t \rightarrow b H^{+}$ is kinematically possible and therefore,  type (2) diagrams may develop an absorptive part.   These scenarios are called  (\textbf{i}) in our paper and we fix for them the mass of the charged scalar, $m_{H^+}$, to be $150$~GeV.  For the other cases, where $m_{H^+}>m_t$,  we take  $m_{H^+}=320$~GeV, as shown in table \ref{escenariosmasas}.  In addition, for a CP conserving scalar potential \cite{Gunion:1989we} we have to impose that $m_{h} \leq m_{H}$. We  define four different  mass scenarios:  two with three  light neutral 
scalars (\textbf{I} and \textbf{Ii}) and  two 
with $h$ as the only light scalar (\textbf{II} and \textbf{IIi}). The other possible two, with the CP-odd scalar $A$ being the lightest one, are disfavored by present LHC data
 \cite{Khachatryan:2015baw,Celis:2013rcs} and are not considered here. 

\begin{table}[H]
\begin{center}
\begin{tabular}{|c|c|c|c|c|c|}
\hline
\multicolumn{5}{|c|}{Scalar mass scenarios (in GeV)}&\multirow{2}{*}{Type of line and color}\\ 
\cline{1-5} 
& $m_h$&$m_H$&$m_A$&$m_{H^\pm}$&\\
\hline
I & $125.09$& $173.21$& $150$ & $320$ & \includegraphics[width=2.5cm,height=0.3cm]{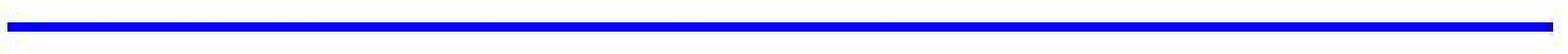} \\
\hline 
Ii & $125.09$ & $173.21$& $150$ & $150$ &  \includegraphics[width=2.5cm,height=0.3cm]{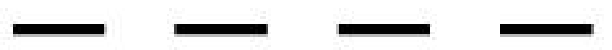} \\
\hline 
II & $125.09$ & $866.05$ & $866.05$& $320$ & \includegraphics[width=2.5cm,height=0.3cm]{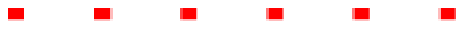}\\
\hline 
IIi & $125.09$ & $866.05$ & $866.05$& $150$ & \includegraphics[width=2.5cm,height=0.3cm]{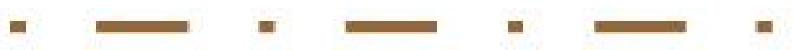}\\
\hline 
\end{tabular}
\caption{Different scalar mass scenarios taken for the analysis.  As specified in the table, each scenario is identified by a different color and type of line in the plots.}
\label{escenariosmasas}
\end{center}
\end{table}
The set of scenarios given in table \ref{escenariosmasas} allows us to investigate the whole meaningful parameter space and  to determine the regions  where $V_R$ strongly differs  from the SM-EW prediction. In all scenarios the value of the heaviest (scalar or pseudoscalar particle) mass, $866.05$~GeV, is fixed by setting $r_{heaviest}=(m_{heaviest})/m_t=5$.

For our numerical analysis we define $Q^{\rm Im}_V$ as the ratio of the  imaginary part of the $V_R$ coupling in the $A2HDM$ to the SM-EW:

\begin{equation}\label{eq:qim}
Q_V^{\rm Im}\equiv \frac{Im\left(V^{A2HDM}_{R}\right)}{Im\left(V^{EW}_{R}\right)}.
\end{equation}
 
Regarding the analysis of the $V_R$ real part, due to the uncertainty already commented in the SM-EW, we present  the results for the {\it A2HDM}  in terms  of  $Re\left(V_R\right) = V_R^{\rm Re}$.

For the four different mass scenarios defined in table \ref{escenariosmasas},  we study  the $V_R$ dependence on the four alignment parameters $\rho_{d, u}$, $\theta_{u,d}$, and on the scalar mixing angle $\gamma$. 
We  show the results for conservative values of the modulus, {\it i.e.} for $\rho_{u,d} \sim 1$. Larger values of these modulus will certainly produce large deviations from the SM predictions but these values are disfavoured with present data \cite{Jung:2012vu,Jung:2010ik}.

\begin{figure}[H] 
\begin{minipage}[b]{.49\linewidth}
\centering\includegraphics[width=75mm]{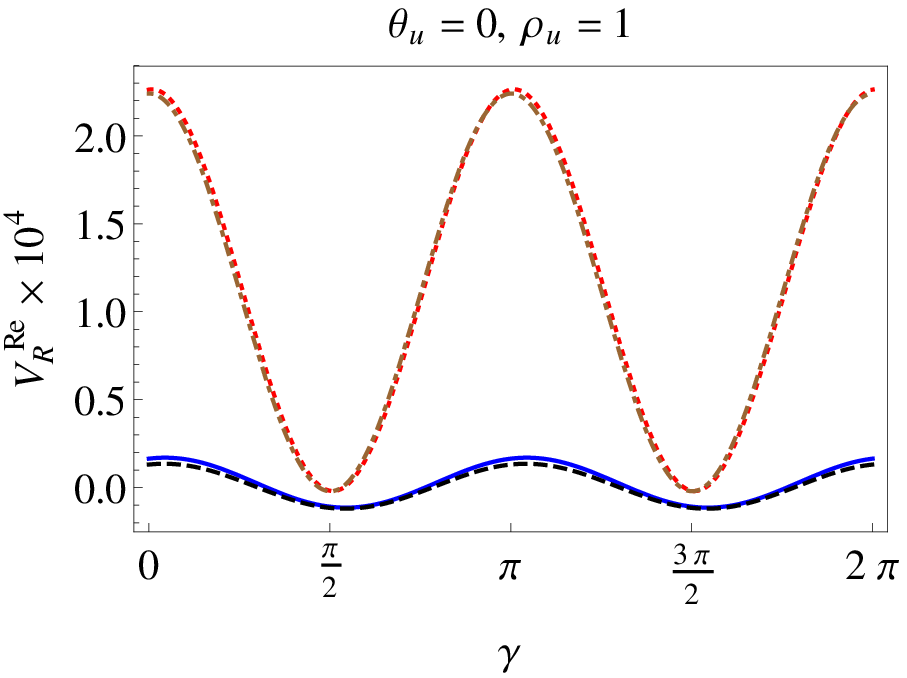}
\end{minipage}
\begin{minipage}[b]{.49\linewidth}
\centering\includegraphics[width=75mm]{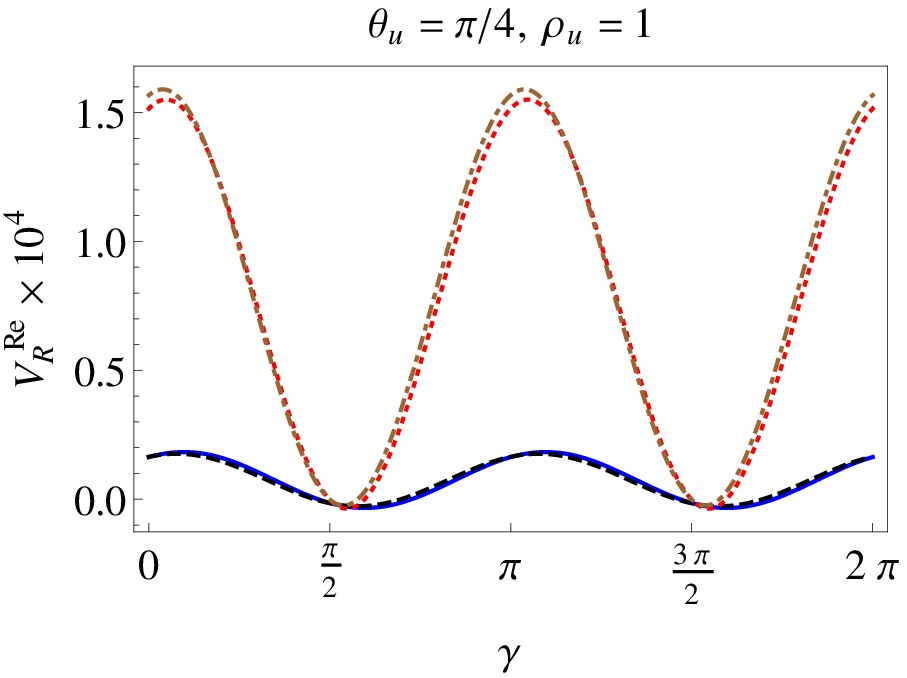}
\end{minipage}
\centering\includegraphics[width=75mm]{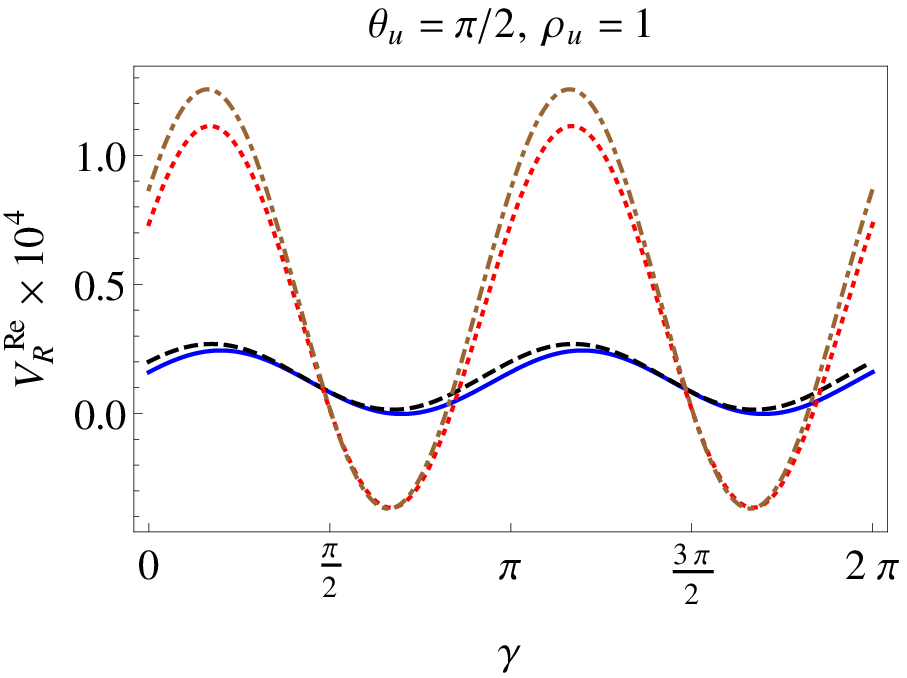}
\caption{$V_R^{\rm Re}=Re\left(V_R^{A2HDM}\right)$, as a function of the $\gamma$ scalar mixing angle, for different $\theta_u$ values and  $p_{u,d}=1$, $\theta_d=\pi/4$.}
\label{plot1}
 \end{figure}

In figure \ref{plot1} we show the dependence of  $V_R^{\rm Re}$ on the $\gamma$ mixing angle, for different values of the $\theta_u$ parameter, with $\rho_{u,d}=1$ and fixing $\theta_d=\pi/4$. $V_R^{\rm Re}$  in the {\it A2HDM} can be three orders of magnitude bigger than the  SM-EW prediction for  scenarios II and IIi, while it can be  one order of magnitude larger for scenarios I and Ii. The behaviour with the $\gamma$ parameter always exhibits the usual  oscillating dependence.   We checked that these results do not depend crucially on the particular $\theta_d$ value chosen. Similar values -with a slight shift of the central values of the $V_R$ coupling- are found when fixing $\theta_u=\pi/4$ and varying $\theta_d$.

\begin{figure}[H] 
\begin{minipage}[b]{.49\linewidth}
\centering\includegraphics[width=75mm]{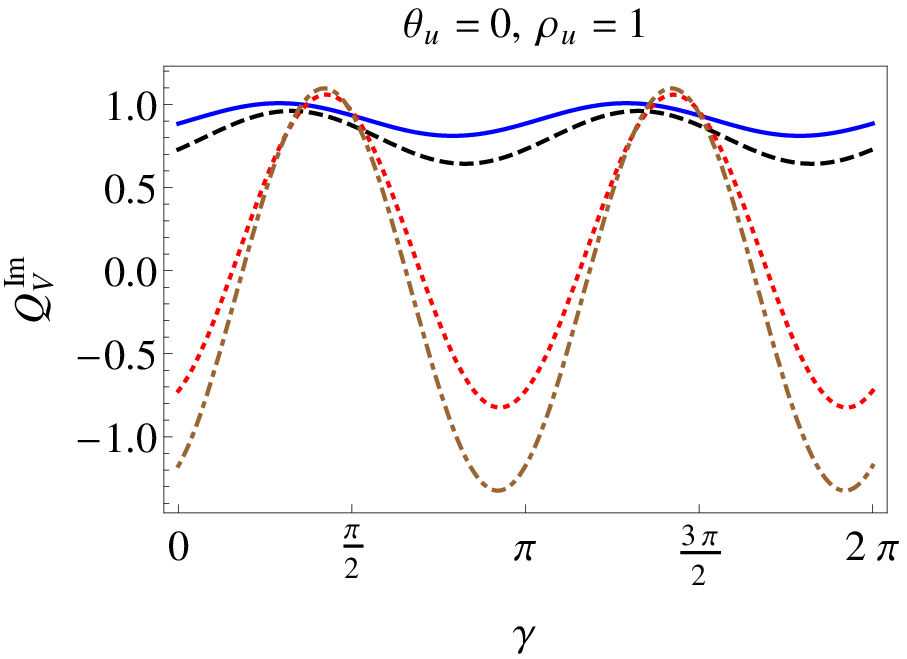}
\end{minipage}
\begin{minipage}[b]{.49\linewidth}
\centering\includegraphics[width=75mm]{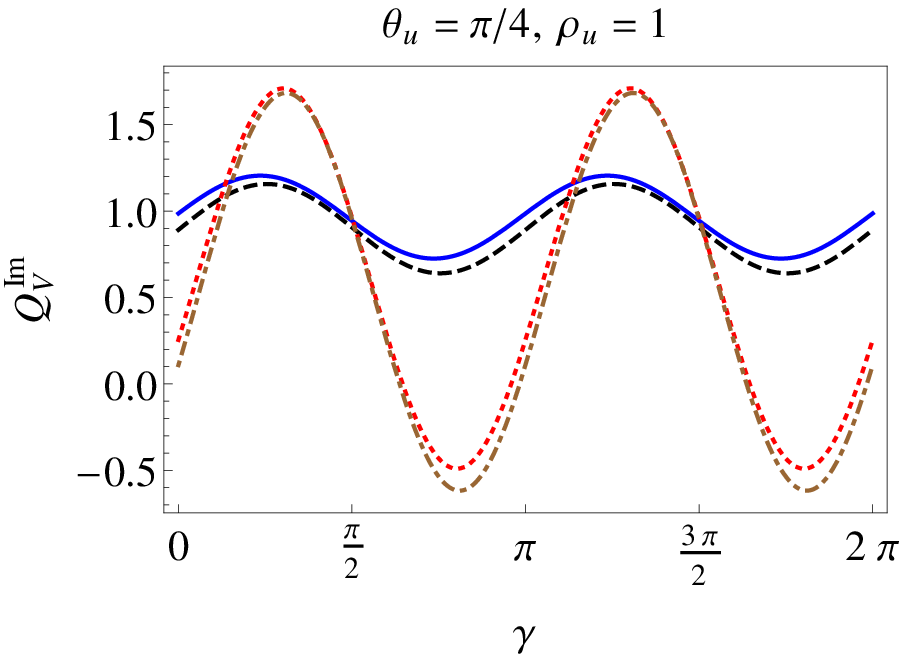}
\end{minipage}
\centering\includegraphics[width=75mm]{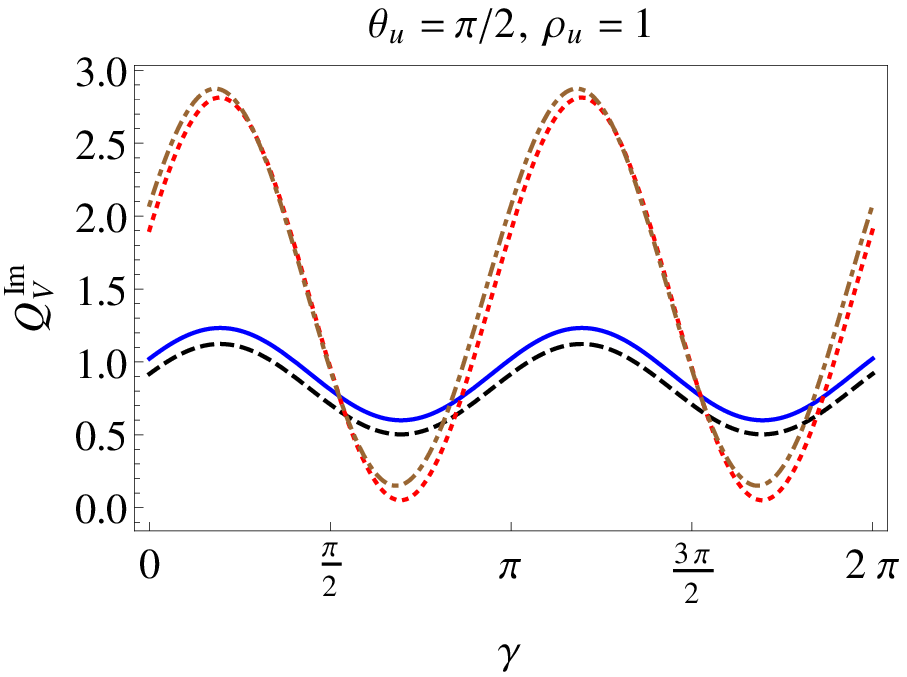}
\caption{$Q_{V}^{\rm Im}$, eq. (\ref{eq:qim}) dependence  with the $\gamma$ scalar mixing angle, for different  $\theta_u$ values and for $\rho_{u,d}=1$, $\theta_d=\pi/4$.}
\label{plot2}
 \end{figure}

In figure \ref{plot2} we show the behaviour of $Q_V^{\rm Im}$  for the same set of parameters as given in figure \ref{plot1}. For  $\rho_{u,d}=1$, $\theta_d=\pi/4$, and $\theta_u$ given in the plots, it can be up to three times larger than the SM-EW value, as can be seen in the third plot of figure \ref{plot2} (scenarios II and IIi). For scenarios I and Ii, the deviation from the SM-EW value is much smaller. The figures  show the expected dependence of the observable with $\gamma$ as a combination of $\sin\gamma$ and $\cos\gamma$. As in the real part of the coupling, the plots for $\mathrm{Im}(V_R)$ present similar behaviour as the one shown in figure \ref{plot2}, with a small shift of their central values, when interchanging $\theta_u\leftrightarrow\theta_d$.

The $V_R$ coupling is more sensitive to the values of $\rho_u$ than  to those of $\rho_d$. The last one may move over a wide range of values ($1<\rho_d<10$) without changing crucially the results. 
In the following we fix the values of the $\rho_d=1$ and $\theta_d=\pi/4$ as a representative choice of these parameters and we study  the dependence of $V_R$ with the rest of the parameters of the model. 

\begin{figure}[H] 
\begin{minipage}[b]{.49\linewidth}
\centering\includegraphics[width=75mm]{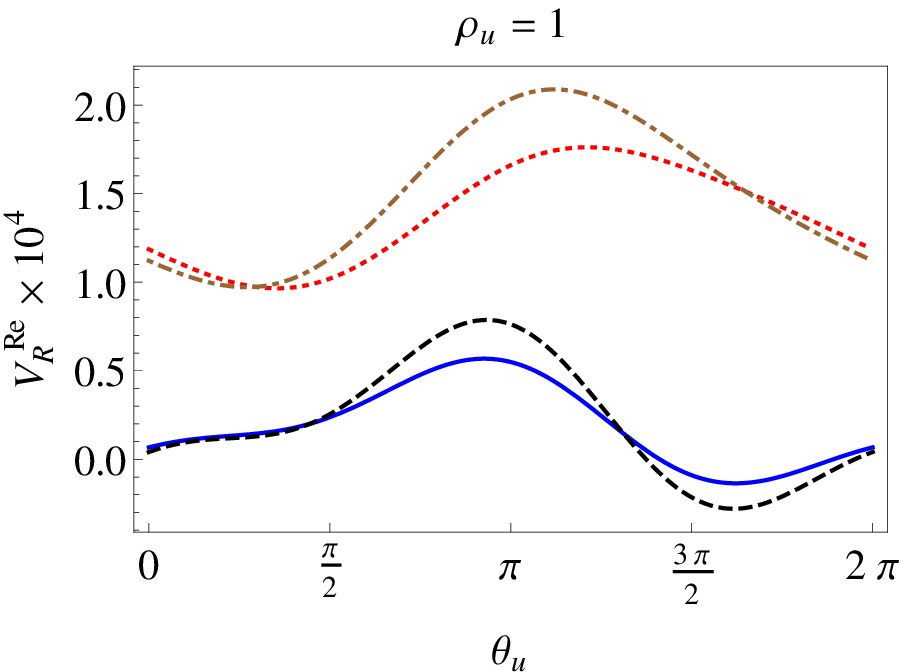}
\end{minipage}
\begin{minipage}[b]{.49\linewidth}
\centering\includegraphics[width=75mm]{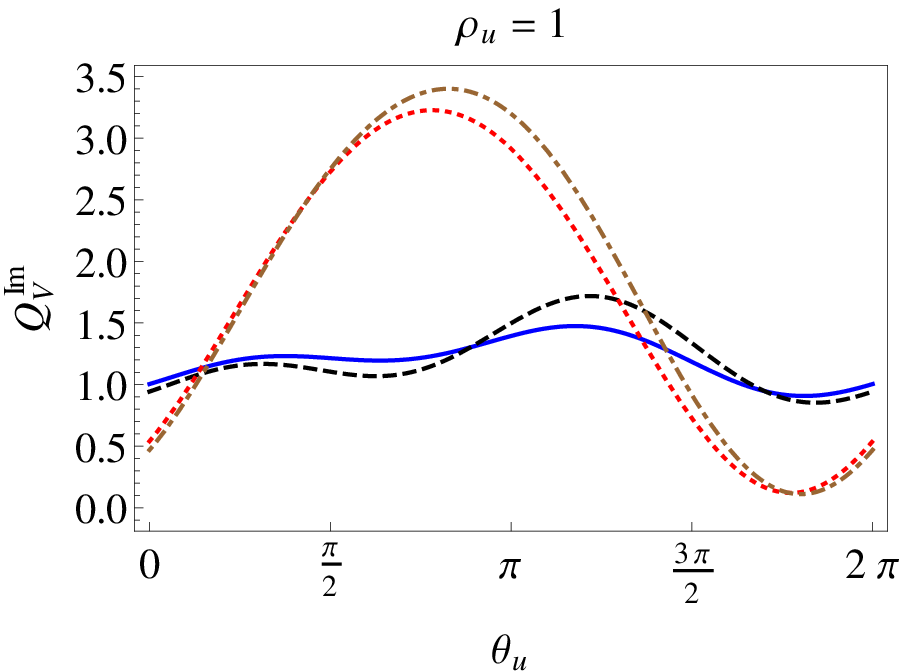}
\end{minipage}
\caption{$V_R^{\rm Re}=Re(V_R^{A2HDM})$ and $Q_{V}^{\rm Im}$ (eq. (\ref{eq:qim})) as function of the $\theta_u$ parameter, with $\gamma=\pi/4$, $\rho_d=1$ and $\theta_d=\pi/4$.
}\label{plot3}
\end{figure}
 
\begin{figure}[H]
\begin{minipage}[b]{.495\linewidth}
\centering\includegraphics[width=77mm]{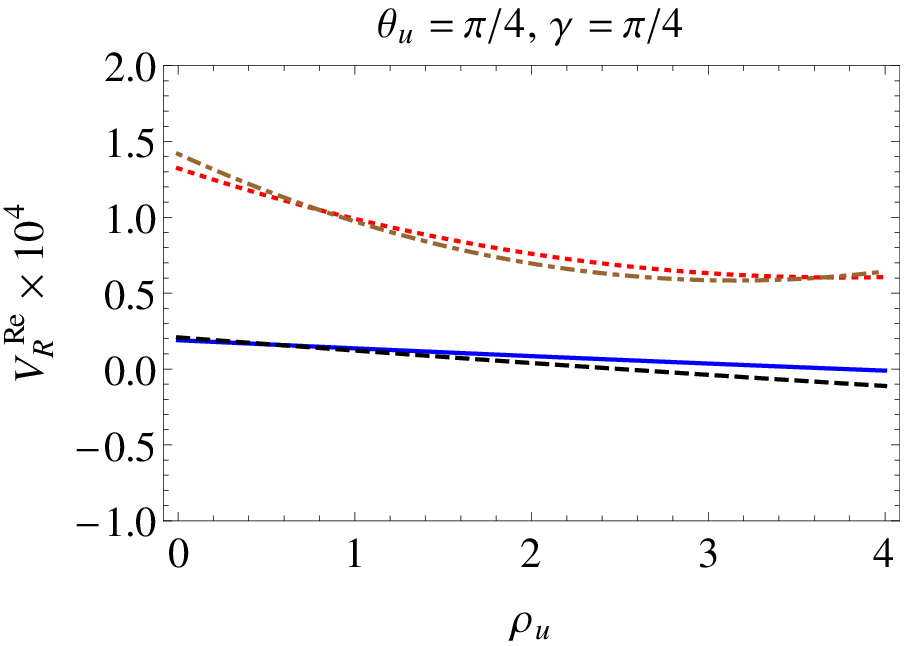}
\end{minipage}
\begin{minipage}[b]{.495\linewidth}
\centering\includegraphics[width=74mm]{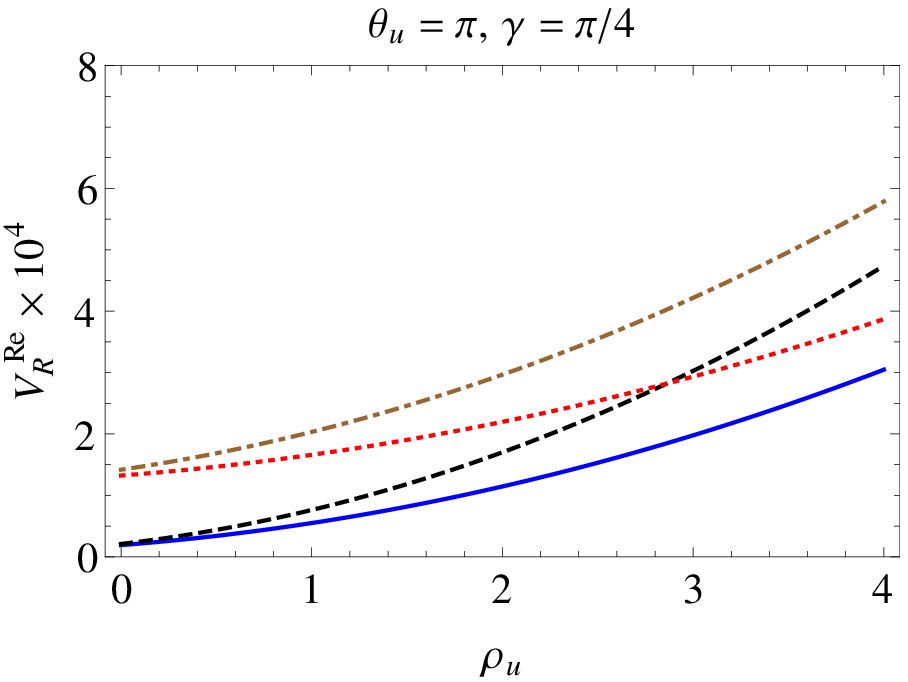}
\end{minipage}
\begin{minipage}[b]{.495\linewidth}
\centering\includegraphics[width=76mm]{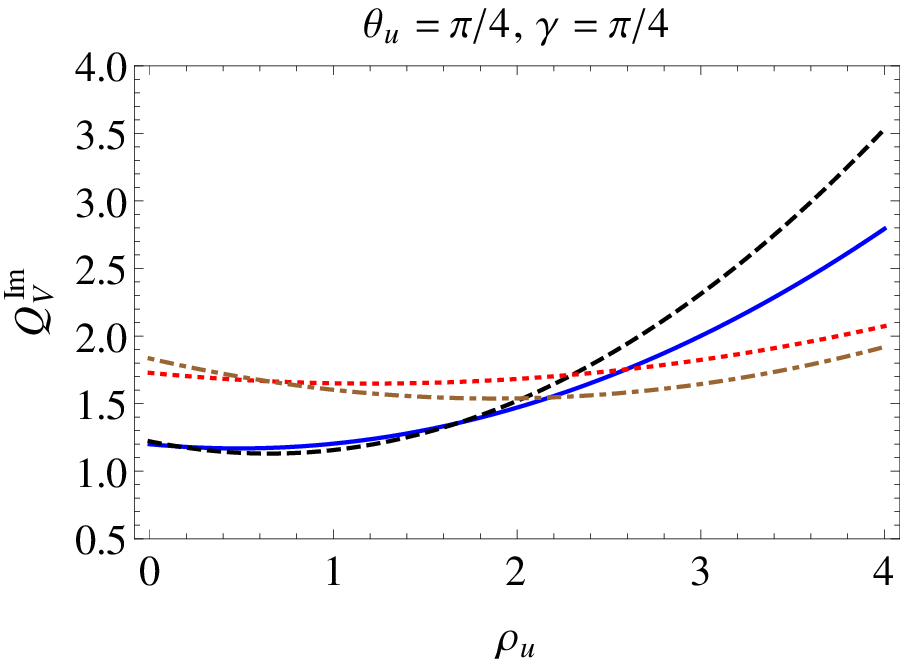}
\end{minipage}
\begin{minipage}[b]{.495\linewidth}
\centering\includegraphics[width=74mm]{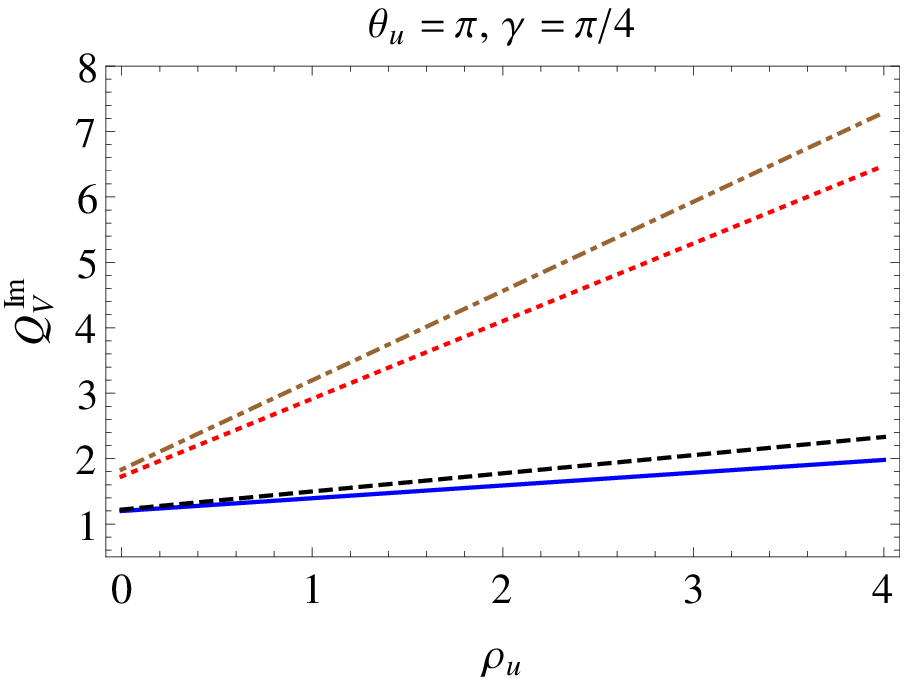}
\end{minipage}
\caption{$V_R^{\rm Re}=Re(V_R^{A2HDM})$ and $Q_{V}^{\rm Im}$, eq. (\ref{eq:qim}),  as a function of  $\rho_u$ for values of $\theta_u$ given in the plots and for fixed values $\gamma=\pi/4$, $\rho_d=1$ and $\theta_d=\pi/4$.}
\label{plot4}
 \end{figure}
 
In figure \ref{plot3} we show $V_R$ (real and imaginary parts) as functions of the $\theta_u$ angle, for the scalar mixing angle $\gamma = \pi/4$. As seen there, the real part can be three (two)  orders of magnitude bigger that the SM-EW one for scenarios II and IIi (I and Ii), while the imaginary part can take values up to three times larger than  the SM-EW prediction, for scenarios II and IIi.

In figure \ref{plot4} we present the dependence of  $V_R$ with the coupling parameter $\rho_u$. The plots show that $V_R^{\rm Re}$ is three (two) orders of magnitude larger than the SM-EW value, for scenarios II and IIi (I and Ii). Besides, for large values of the $\rho_u$ parameter, $V_R^{\rm Re}$ grows with $\rho_u$ independently of the values of the other  parameters of the model, such as $\gamma$ and $\theta_u$. A similar behaviour is found for the imaginary part of  $V_R$, that can be a factor seven larger than the SM-EW one for large values of $\rho_u$.
 
Finally, we compute $V_R$ for Type-I \cite{Haber:1978jt, Hall:1981bc} and Type-II \cite{Hall:1981bc,Donoghue:1978cj} {\it 2HDM} $\!$\footnote{See ref. \cite{Arhrib:2016vts} for a study of the values of the different top couplings in Type-I and Type-II {\it 2HDM}.}.  
In table \ref{tabla3} we show the  $\varsigma_{u,d}$ values that reproduce the Type-I and Type-II models. These models have a discrete ${\cal{Z}}_2$ symmetry in order to avoid  tree level FCNC. 

\begin{table}[ht]
\begin{center}
\begin{tabular}{|c|c|c|}
\hline
Model & $\varsigma_d$ & $\varsigma_u$ 
\\ \hline
Type-I & $\cot{\beta}$ & $\cot{\beta}$ 
\\ \hline
Type-II & $-\tan{\beta}$ & $\cot{\beta}$ 
\\ \hline
\end{tabular}
\end{center}
\caption{Values for $\varsigma_{u,d}$ that reproduce the Type-I and Type-II {\it 2HDM}.}\label{tabla3}
\end{table}

 \begin{figure}[H] 
\begin{minipage}[b]{.49\linewidth}
\centering\includegraphics[width=77mm]{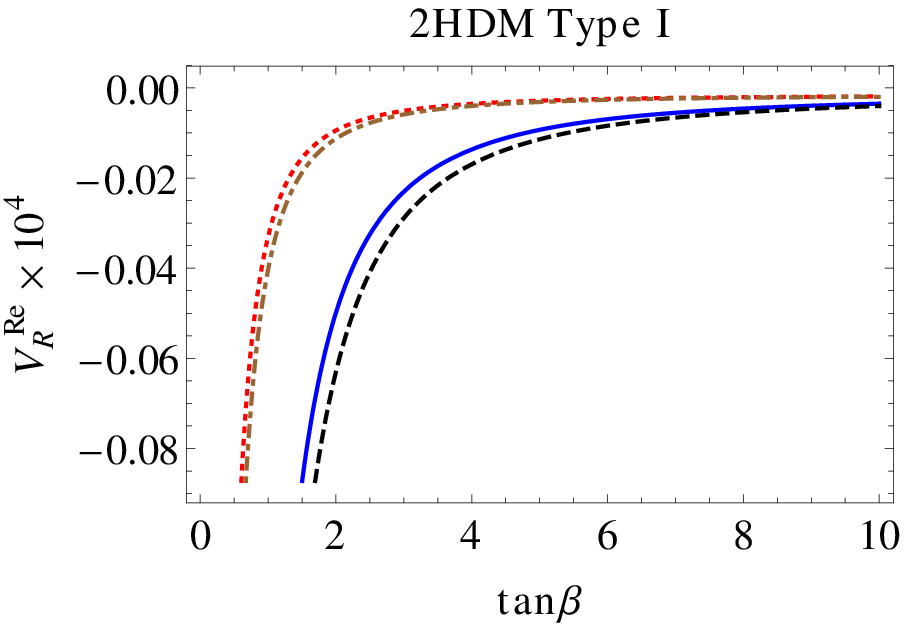}
\end{minipage}
\begin{minipage}[b]{.49\linewidth}
\centering\includegraphics[width=74mm]{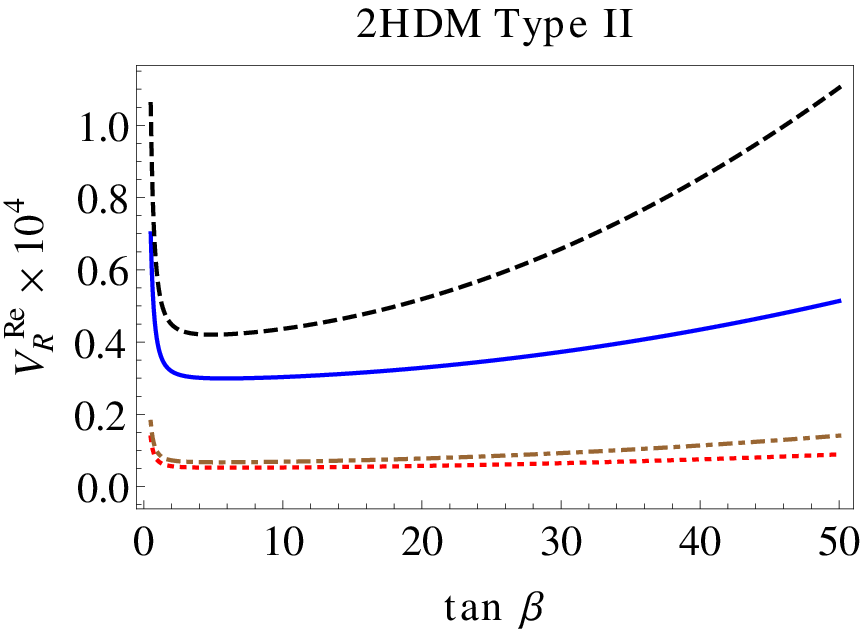}
\end{minipage}
\caption{Type-I {\it 2HDM} prediction for  $V_R^{Re}$  (left) and $Q_V^{Im}$ (right)  as a function of $\tan\beta$, with $\gamma=-\pi/2$.}
\label{plot5}
 \end{figure}

 For Type-I and Type-II {\it 2HDM}, we present the results as a function of $\tan\beta$, for the different mass scenarios considered. We work on the alignment limit, $\gamma=-\pi/2$, where the neutral scalar $h$ has SM-like couplings to the photon and to the weak bosons. The results for the real part of $V_R$ are shown in figure \ref{plot5}.  For Type-I model, $V_R$ takes values one order of magnitude larger than the SM-EW one, for  $1 < \tan\beta <4$ and for all mass scenarios;  for $\tan\beta>>4$  it approaches  the SM-EW value. Note that  for Type-I {\it 2HDM} the Yukawa couplings go to zero in the large  $\tan\beta$ limit.   For Type-II model the value of $V_R$ grows with $\tan\beta$, reaching values close to $10^{-4}$ ($10^{-5}$) for $\tan \beta \simeq 50$ in the   mass scenarios I and Ii (II and IIi). We also find that, within these models,  $V_R^{\rm Im}$ is very close to the SM-EW value and almost constant  for the  considered  mass scenarios.
 
\section{Conclusions}

We  computed the one-loop contribution to $V_R$ in the {\it A2HDM}. In the SM, accidental cancellation among the one loop EW parts   results in  values for  $V_R^{\rm Re}$ two orders of magnitude smaller than expected from each diagram. This cancellation does not take place in the  {\it A2HDM}. Then, depending on the values of the parameters of the model, the magnitude of $V_R^{\rm Re}$  can be  three orders of magnitude larger than the SM-EW prediction (i.e. close to  $10^{-4}$) and close to the leading QCD contribution. $V_R^{\rm Im}$  can  be one order of magnitude  larger than the SM prediction for  $\rho_{u,d}>4$ but, for $\rho_{u,d}\sim 1$, its magnitude is only a few times larger. For Type-II (Type-I) {\it 2HDM},  $V_R^{\rm Re}$  can grow up to two (one) orders of magnitude with respect the SM-EW value, for $\tan\beta\approx 10$ and depending of the mass scenarios considered, while the imaginary part remains basically of the same order as in the SM-EW. As it is shown in our  previous work \cite{Gonzalez-Sprinberg:2015dea}, new observables for the LHC and next generation colliders can provide a direct measurement of the right top coupling $V_R$. 

\section{Acknowledgments}

This work has been supported, in part, by the Ministerio de Economía y Competitividad (MINECO), Spain, under grants FPA2014-54459-P and SEV-2014-0398; by Generalitat Valenciana, Spain, under grant PROMETEOII2014-087. G.A.G-S. acknowledges the support of CSIC and Pedeciba, Uruguay. C.A. acknowledges the support by the Spanish Government and ERDF funds from the EU Commission [Grant No. FPA2014-53631-C2-1-P] and by CONICYT Fellowship “Becas Chile” Grant No. 74150052. R.M. also thanks to COLCIENCIAS.

 \begin{appendices}

\section{{\it A2HDM} contribution to \texorpdfstring{$V_R$}{Vr}}
\label{app:AppendixA}

Following the notation of ref. \cite{Gonzalez-Sprinberg:2015dea}, we define
\begin{eqnarray}
A_a&=& x^2 \left[\left((y-1)  r_b^2+1\right) y-  r_w^2 (y-1)\right]-r_a^2 (x-1), \\ 
\widetilde{A}_a&=& A_a\left(r_w\rightarrow r_{H^+}\right),\\
B_a &=& x \left\{\left[\left(x (y-1)+1\right)  r_b^2+x-1\right]y -r_a^2 (y-1)\right\}\nonumber\\
&&\hspace*{5cm} - r_w^2 (x-1) \left[x (y-1)+1\right], \\ 
\widetilde{B}_a&=& B_a\left(r_w\rightarrow r_{H^+}\right),\\ 
C_a &=& (x-1) (x y-1)  r_b^2-r_w^2 (x-1) x (y-1)\ +r_a^2 xy+x (y-1) (x y-1),\label{anal:tgw}
\end{eqnarray}
with
\begin{equation}
r_x\equiv \frac{m_x}{m_t},
\label{rx}
\end{equation}
and
\begin{equation}
\begin{array}{lcl}
 y^{H}_{d}&=& \cos\gamma +\sin\gamma \, \varsigma_{d}\, ,  \\
   y^{h}_{d}&=& -\sin\gamma +\cos\gamma\, \varsigma_{d}\, ,  \\ 
   y^{A}_{d}&=& i \, \varsigma_{d}\, ,
\end{array}
\qquad\quad
\begin{array}{lcl}
  y^{H}_{u}&=& \cos\gamma +\sin\gamma \, \varsigma^*_{u}\, ,\\
   y^{h}_{u}&=& -\sin\gamma +\cos\gamma\, \varsigma^*_{u}\, ,\\
   y^{A}_{u}&=& -i \, \varsigma^*_{u}\, .
\end{array}
\end{equation}
Then, we have the following expressions for the new contributions, listed in table \ref{tabla2}:

\noindent - Type (1) diagrams.
\bea
\lefteqn{I^{tHH^-}+I^{thH^-}+I^{tAH^-}=\frac{1}{16\pi s_w^2 r_w^2}\times}
\nonumber\\
&&
\int_0^1 dx \int_0^1 dy\, x\left\{ \varsigma_d\left[y_u^H \sin\gamma\, \ln\frac{\widetilde{A}_H}{\widetilde{A}_A}+y_u^h \cos\gamma\; \ln\frac{
\widetilde{A}_h}{\widetilde{A}_A} \right]
\right.
\nonumber\\
&&
\left.
+y x \left[\frac{v_R^{uHH^+}}{\widetilde{A}_H}\sin\gamma
+\frac{v_R^{uhH^-}}{\widetilde{A}_h}\cos\gamma 
-i\frac{v_R^{uAH^-}}{\widetilde{A}_A} \right]  \right\},
\eea
with
\bea
v_R^{u\varphi_iH^-}&=&-y_u^{\varphi_i}\left\{\varsigma_d\left[r_b^2(1-y)x+(1-x)\right]-\varsigma_u\right\}\nonumber\\
&& +y_u^{\varphi_i^*}\left[\varsigma_u(1-xy)-\varsigma_d\right],\quad \varphi_i=h,\, H,\, A.
\eea

\noindent - Type (2) diagrams.
\bea
\lefteqn{I^{bH^+H}+I^{bH^+h}+I^{bH^+A}=-\frac{1}{16\pi s_w^2 r_w^2}\times} 
\nonumber\\
&&
\int_0^1 dx \int_0^1 dy\, x \left\{ \varsigma_u\left[ y_d^{H^*}\sin\gamma\, \ln\frac{\widetilde{B}_H}{\widetilde{B}_A}+ y_d^{h^*} \cos\gamma\, 
\ln\frac{\widetilde{B}_h}{\widetilde{B}_A} \right]
\right.
\nonumber\\
&&
\left.
+y x \left[\frac{v_R^{dH^+H}}{\widetilde{B}_H}\sin\gamma 
+\frac{v_R^{dH^+h}}{\widetilde{B}_h}\cos\gamma
-i\frac{v_R^{dH^+A}}{\widetilde{B}_A} \right]  \right\},
\eea
with
\bea
v_R^{dH^+\varphi_i}&=& -y_d^{\varphi_i}\, r_b^2\left[\varsigma_d(1-xy)-\varsigma_u\right]\nonumber\\
&&+y_d^{\varphi_i^*}\left\{\varsigma_u\left[(1-y)x\, r_b^2+(1-x)\right]-\varsigma_d r_b^2\right\},\quad \varphi_i=h,\, H,\, A.
\eea

\noindent - Type (3) diagrams.
\bea
\lefteqn{I^{G^0tb}+I^{H tb}+I^{h tb}+I^{A tb}=\frac{1}{16\pi s_w^2}\times} 
\nonumber\\
&&
\int_0^1 dx \int_0^1 dy\, x \left\{ 
x(1-x)(1-y)\left[-\frac{1}{C_Z}+\frac{y_d^{H^*}y_u^H}{C_H} +\frac{y_d^{h^*}y_u^h}{C_h}+\frac{y_d^{A^*}y_u^A}{C_A} \right]\right. \nonumber\\
&&
\left.-\frac{1}{r_w^2}\left[y_d^{H^*}y_u^H\, \ln\frac{C_H}{C_Z}+y_d^{h^*}y_u^h\, 
\ln\frac{C_h}{C_Z}+y_d^{A^*}y_u^A\, \ln\frac{C_A}{C_Z} \right] 
 \right\}.
\eea

\noindent - Type (6) diagrams.
\bea
\lefteqn{I^{tG^0 G^-}+I^{tHG^-}+I^{thG^-}=\frac{1}{16\pi s_w^2 r_w^2}\times} 
\nonumber\\
&&
\int_0^1 dx \int_0^1 dy\, x \left\{ x^2y\left[
-\frac{1+y-r_b^2(1-y)}{A_Z}+\frac{(r_b^2(y-1)+1)y_u^H-y\ y_u^{H^*}}{A_H}
\cos\gamma\right.
\right.
\nonumber\\
&&
 \left.-\frac{(r_b^2(y-1)+1)y_u^h-y\ y_u^{h^*}}{A_h}\sin\gamma\right]\nonumber\\
&&
+ \left.y_u^H\cos\gamma\,\ln\frac{A_H}{A_Z}-y_u^h\sin\gamma\, \ln\frac{A_h}{A_Z}  \right\}.
\eea

\noindent - Type (7) diagrams.

\bea
\lefteqn{I^{bG^+G^0}+I^{b G^+H}+I^{b G^+h}=\frac{1}{16\pi s_w^2 r_w^2}\times} 
\nonumber\\
&&
\int_0^1 dx \int_0^1 dy\, x \left\{ xy\left[
\frac{1-x-r_b^2(1-x(1-2y))}{B_Z}
\right.\right.
\nonumber\\
&&
+\frac{(r_b^2((y-1)x+1)+x-1)y_d^{H^*}-x y\ r_b^2 y_d^H}{B_H}\cos\gamma
\nonumber\\
&&
\left.-\frac{(r_b^2((y-1)x+1)+x-1)y_d^{h^*}-x y\ r_b^2 y_d^h}{B_h}\sin\gamma
\right]
\nonumber\\
&&
\left. +y_d^{H^*}\cos\gamma\, \ln\frac{B_H}{B_Z}-y_d^{h^*}\sin\gamma\, \ln\frac{B_h}{B_Z} 
\right\}.
\eea

The contribution from type (4), $\left\{t\varphi_iW\right\}$, and type (5), $\left\{bW\varphi_i\right\}$, diagrams ($\varphi_i=h,\, H$) is zero as in the SM:
\be
V_R^{t\varphi_iW^-}=V_R^{bW^+\varphi_i}=0, \quad \varphi_i=h,\, H.
\ee

Notice that in the limit $\varsigma_{u,d}\rightarrow 0$, and fixing $\gamma=-\pi/2$ to identifying $h$ with the standard Higgs, we recover the SM result \cite{Gonzalez-Sprinberg:2015dea}.
\end{appendices}
\bibliographystyle{JHEP}
\bibliography{BibLu_mod}

\providecommand{\href}[2]{#2}\begingroup\raggedright\begin{thebibliography}{10}

\bibitem{Selvaggi:2015sdf}
M.~Selvaggi, {\it {Perspectives for Top quark physics at High-Luminosity LHC}},
   {\em PoS} {\bf TOP2015} (2016) 054,
  [\href{http://xxx.lanl.gov/abs/1512.04807}{{\tt arXiv:1512.04807}}].

\bibitem{Barletta:2013ooa}
W.~Barletta, M.~Battaglia, M.~Klute, M.~Mangano, S.~Prestemon, L.~Rossi, and
  P.~Skands, {\it {Working Group Report: Hadron Colliders}},  in {\em
  {Proceedings, Community Summer Study 2013: Snowmass on the Mississippi
  (CSS2013): Minneapolis, MN, USA, July 29-August 6, 2013}}, 2013.
\newblock \href{http://xxx.lanl.gov/abs/1310.0290}{{\tt arXiv:1310.0290}}.

\bibitem{Abe:1995hr}
{\bf CDF Collaboration} Collaboration, F.~Abe et~al., {\it {Observation of top
  quark production in $\bar{p}p$ collisions}},  {\em Phys. Rev. Lett.} {\bf 74}
  (1995) 2626--2631, [\href{http://xxx.lanl.gov/abs/hep-ex/9503002}{{\tt
  hep-ex/9503002}}].

\bibitem{Abachi:1995iq}
{\bf D0 Collaboration} Collaboration, S.~Abachi et~al., {\it {Observation of
  the top quark}},  {\em Phys. Rev. Lett.} {\bf 74} (1995) 2632--2637,
  [\href{http://xxx.lanl.gov/abs/hep-ex/9503003}{{\tt hep-ex/9503003}}].

\bibitem{Deterre:2012vn}
C.~Deterre, {\it {$W$ helicity and constraints on the $Wtb$ vertex at the
  Tevatron}},  {\em Nuovo Cim.} {\bf C035N3} (2012) 125--129,
  [\href{http://xxx.lanl.gov/abs/1203.6802}{{\tt arXiv:1203.6802}}].

\bibitem{Abazov:2011pm}
{\bf D0 Collaboration} Collaboration, V.~M. Abazov et~al., {\it {Search for
  anomalous $Wtb$ couplings in single top quark production in $p\bar{p}$
  collisions at $\sqrt{s} = 1.96$ TeV}},  {\em Phys. Lett. B} {\bf 708} (2012)
  21--26, [\href{http://xxx.lanl.gov/abs/1110.4592}{{\tt arXiv:1110.4592}}].

\bibitem{Abazov:2008sz}
{\bf D0} Collaboration, V.~M. Abazov et~al., {\it {Search for anomalous Wtb
  couplings in single top quark production}},  {\em Phys. Rev. Lett.} {\bf 101}
  (2008) 221801, [\href{http://xxx.lanl.gov/abs/0807.1692}{{\tt
  arXiv:0807.1692}}].

\bibitem{Bernreuther:2008ju}
W.~Bernreuther, {\it {Top quark physics at the LHC}},  {\em J. Phys. G} {\bf
  35} (2008) 083001, [\href{http://xxx.lanl.gov/abs/0805.1333}{{\tt
  arXiv:0805.1333}}].

\bibitem{Bardhan:2016txk}
D.~Bardhan, G.~Bhattacharyya, D.~Ghosh, M.~Patra, and S.~Raychaudhuri, {\it
  {Detailed analysis of flavor-changing decays of top quarks as a probe of new
  physics at the LHC}},  {\em Phys. Rev.} {\bf D94} (2016), no.~1 015026,
  [\href{http://xxx.lanl.gov/abs/1601.04165}{{\tt arXiv:1601.04165}}].

\bibitem{Bernreuther:2015yna}
W.~Bernreuther, D.~Heisler, and Z.-G. Si, {\it {A set of top quark spin
  correlation and polarization observables for the LHC: Standard Model
  predictions and new physics contributions}},  {\em JHEP} {\bf 12} (2015) 026,
  [\href{http://xxx.lanl.gov/abs/1508.05271}{{\tt arXiv:1508.05271}}].

\bibitem{Schilling:2012dx}
F.-P. Schilling, {\it {Top Quark Physics at the LHC: A Review of the First Two
  Years}},  {\em Int. J. Mod. Phys.} {\bf A27} (2012) 1230016,
  [\href{http://xxx.lanl.gov/abs/1206.4484}{{\tt arXiv:1206.4484}}].

\bibitem{Bernardo:2014vha}
C.~Bernardo, N.~F. Castro, M.~C.~N. Fiolhais, H.~Gonçalves, A.~G.~C. Guerra,
  M.~Oliveira, and A.~Onofre, {\it {Studying the $Wtb$ vertex structure using
  recent LHC results}},  {\em Phys. Rev.} {\bf D90} (2014), no.~11 113007,
  [\href{http://xxx.lanl.gov/abs/1408.7063}{{\tt arXiv:1408.7063}}].

\bibitem{Hawkings:2015ega}
R.~Hawkings, {\it {Top quark physics at the LHC}},  {\em Comptes Rendus
  Physique} {\bf 16} (2015) 424--434.

\bibitem{Cristinziani:2016vif}
M.~Cristinziani and M.~Mulders, {\it {Top-quark physics at the Large Hadron
  Collider}},  \href{http://xxx.lanl.gov/abs/1606.00327}{{\tt
  arXiv:1606.00327}}.

\bibitem{Calvet:2014fxa}
{\bf ATLAS} Collaboration, D.~Calvet, {\it {Search for New Physics with Top
  quarks in ATLAS at 8 TeV ($t\bar{b}$, $t\bar{t}$, vector-like quarks)}},  in
  {\em {Proceedings, 20th International Conference on Particles and Nuclei
  (PANIC 14): Hamburg, Germany, August 24-29, 2014}}, pp.~579--582, 2014.

\bibitem{Pagano:2013goa}
{\bf ATLAS, CMS} Collaboration, D.~Pagano, {\it {Measurements of new physics in
  top quark decay at LHC}},  {\em J. Phys. Conf. Ser.} {\bf 452} (2013), no.~1
  012011.

\bibitem{Gaitan:2015aia}
R.~Gaitan, E.~A. Garces, J.~H.~M. de~Oca, and R.~Martinez, {\it {Top quark
  Chromoelectric and Chromomagnetic Dipole Moments in a Two Higgs Doublet Model
  with CP violation}},  {\em Phys. Rev.} {\bf D92} (2015), no.~9 094025,
  [\href{http://xxx.lanl.gov/abs/1505.04168}{{\tt arXiv:1505.04168}}].

\bibitem{delAguila:2002nf}
F.~del Aguila and J.~Aguilar-Saavedra, {\it {Precise determination of the Wtb
  couplings at CERN LHC}},  {\em Phys.Rev.} {\bf D67} (2003) 014009,
  [\href{http://xxx.lanl.gov/abs/hep-ph/0208171}{{\tt hep-ph/0208171}}].

\bibitem{AguilarSaavedra:2010nx}
J.~Aguilar-Saavedra and J.~Bernab\'eu, {\it {W polarisation beyond helicity
  fractions in top quark decays}},  {\em Nucl. Phys. B} {\bf 840} (2010)
  349--378, [\href{http://xxx.lanl.gov/abs/1005.5382}{{\tt arXiv:1005.5382}}].

\bibitem{Drobnak:2010ej}
J.~Drobnak, S.~Fajfer, and J.~F. Kamenik, {\it {New physics in $t-> b W$ decay
  at next-to-leading order in QCD}},  {\em Phys. Rev.} {\bf D82} (2010) 114008,
  [\href{http://xxx.lanl.gov/abs/1010.2402}{{\tt arXiv:1010.2402}}].

\bibitem{Rindani:2011pk}
S.~D. Rindani and P.~Sharma, {\it {Probing anomalous tbW couplings in
  single-top production using top polarization at the Large Hadron Collider}},
  {\em JHEP} {\bf 1111} (2011) 082,
  [\href{http://xxx.lanl.gov/abs/1107.2597}{{\tt arXiv:1107.2597}}].

\bibitem{Prasath:2014mfa}
A.~V. Prasath, R.~M. Godbole, and S.~D. Rindani, {\it {Longitudinal top
  polarisation measurement and anomalous $Wtb$ coupling}},  {\em Eur. Phys. J.}
  {\bf C75} (2015), no.~9 402, [\href{http://xxx.lanl.gov/abs/1405.1264}{{\tt
  arXiv:1405.1264}}].

\bibitem{Cao:2015doa}
Q.-H. Cao, B.~Yan, J.-H. Yu, and C.~Zhang, {\it {A General Analysis of $Wtb$
  anomalous Couplings}},  \href{http://xxx.lanl.gov/abs/1504.03785}{{\tt
  arXiv:1504.03785}}.

\bibitem{Hioki2016128}
Z.~Hioki and K.~Ohkuma, {\it Full analysis of general non-standard tbw
  couplings},  {\em Physics Letters B} {\bf 752} (2016) 128 -- 130.

\bibitem{Olive:2016xmw}
C.~Patrignani, {\it {Review of Particle Physics}},  {\em Chin. Phys.} {\bf C40}
  (2016), no.~10 100001.

\bibitem{MorenoLlacer:2014tca}
M.~Moreno~Llácer, {\em {Search for CP violation in single top quark events
  with the ATLAS detector at LHC}}.
\newblock PhD thesis, Valencia U., IFIC, 2014.

\bibitem{GonzalezSprinberg:2011kx}
G.~A. Gonz\'alez-Sprinberg, R.~Martinez, and J.~Vidal, {\it {Top quark tensor
  couplings}},  {\em JHEP} {\bf 07} (2011) 094,
  [\href{http://xxx.lanl.gov/abs/1105.5601}{{\tt arXiv:1105.5601}}]. [Erratum:
  JHEP05,117(2013)].

\bibitem{Duarte:2013zfa}
L.~Duarte, G.~A. González-Sprinberg, and J.~Vidal, {\it {Top quark anomalous
  tensor couplings in the two-Higgs-doublet models}},  {\em JHEP} {\bf 1311}
  (2013) 114, [\href{http://xxx.lanl.gov/abs/1308.3652}{{\tt
  arXiv:1308.3652}}].

\bibitem{Bernreuther:2008us}
W.~Bernreuther, P.~Gonzalez, and M.~Wiebusch, {\it {The Top Quark Decay Vertex
  in Standard Model Extensions}},  {\em Eur. Phys. J. C} {\bf 60} (2009)
  197--211, [\href{http://xxx.lanl.gov/abs/0812.1643}{{\tt arXiv:0812.1643}}].

\bibitem{Gonzalez-Sprinberg:2015dea}
G.~A. González-Sprinberg and J.~Vidal, {\it {The top quark right coupling in
  the tbW-vertex}},  {\em Eur. Phys. J.} {\bf C75} (2015), no.~12 615,
  [\href{http://xxx.lanl.gov/abs/1510.02153}{{\tt arXiv:1510.02153}}].

\bibitem{Lampe:1995xb}
B.~Lampe, {\it {Forward - backward asymmetry in top quark semileptonic decay}},
   {\em Nucl.Phys.} {\bf B454} (1995) 506--526.

\bibitem{AguilarSaavedra:2006fy}
J.~Aguilar-Saavedra, J.~Carvalho, N.~F. Castro, F.~Veloso, and A.~Onofre, {\it
  {Probing anomalous Wtb couplings in top pair decays}},  {\em Eur.Phys.J.}
  {\bf C50} (2007) 519--533,
  [\href{http://xxx.lanl.gov/abs/hep-ph/0605190}{{\tt hep-ph/0605190}}].

\bibitem{Grzadkowski:1999iq}
B.~Grzadkowski and Z.~Hioki, {\it {New hints for testing anomalous top quark
  interactions at future linear colliders}},  {\em Phys.Lett.} {\bf B476}
  (2000) 87--94, [\href{http://xxx.lanl.gov/abs/hep-ph/9911505}{{\tt
  hep-ph/9911505}}].

\bibitem{Godbole:2006tq}
R.~M. Godbole, S.~D. Rindani, and R.~K. Singh, {\it {Lepton distribution as a
  probe of new physics in production and decay of the t quark and its
  polarization}},  {\em JHEP} {\bf 0612} (2006) 021,
  [\href{http://xxx.lanl.gov/abs/hep-ph/0605100}{{\tt hep-ph/0605100}}].

\bibitem{Stelzer:1995gc}
T.~Stelzer and S.~Willenbrock, {\it {Spin correlation in top quark production
  at hadron colliders}},  {\em Phys.Lett.} {\bf B374} (1996) 169--172,
  [\href{http://xxx.lanl.gov/abs/hep-ph/9512292}{{\tt hep-ph/9512292}}].

\bibitem{Mahlon:1995zn}
G.~Mahlon and S.~J. Parke, {\it {Angular correlations in top quark pair
  production and decay at hadron colliders}},  {\em Phys.Rev.} {\bf D53} (1996)
  4886--4896, [\href{http://xxx.lanl.gov/abs/hep-ph/9512264}{{\tt
  hep-ph/9512264}}].

\bibitem{Pich:2009sp}
A.~Pich and P.~Tuzon, {\it {Yukawa Alignment in the Two-Higgs-Doublet Model}},
  {\em Phys. Rev. D} {\bf 80} (2009) 091702,
  [\href{http://xxx.lanl.gov/abs/0908.1554}{{\tt arXiv:0908.1554}}].

\bibitem{Lee:1973iz}
T.~D. Lee, {\it {A Theory of Spontaneous T Violation}},  {\em Phys. Rev.} {\bf
  D8} (1973) 1226--1239.

\bibitem{Branco:2011iw}
G.~Branco, P.~Ferreira, L.~Lavoura, M.~Rebelo, M.~Sher, et~al., {\it {Theory
  and phenomenology of two-Higgs-doublet models}},  {\em Phys. Rept.} {\bf 516}
  (2012) 1--102, [\href{http://xxx.lanl.gov/abs/1106.0034}{{\tt
  arXiv:1106.0034}}].

\bibitem{Jung:2010ik}
M.~Jung, A.~Pich, and P.~Tuzon, {\it {Charged-Higgs phenomenology in the
  Aligned two-Higgs-doublet model}},  {\em JHEP} {\bf 1011} (2010) 003,
  [\href{http://xxx.lanl.gov/abs/1006.0470}{{\tt arXiv:1006.0470}}].

\bibitem{Jung:2012vu}
M.~Jung, X.-Q. Li, and A.~Pich, {\it {Exclusive radiative B-meson decays within
  the aligned two-Higgs-doublet model}},  {\em JHEP} {\bf 1210} (2012) 063,
  [\href{http://xxx.lanl.gov/abs/1208.1251}{{\tt arXiv:1208.1251}}].

\bibitem{Chakraborty:2015qja}
{\bf ATLAS, CMS} Collaboration, D.~Chakraborty, {\it {Charged Higgs boson
  searches at the LHC}},  {\em Nucl. Part. Phys. Proc.} {\bf 260} (2015)
  216--220.

\bibitem{Celis:2013ixa}
A.~Celis, V.~Ilisie, and A.~Pich, {\it {Towards a general analysis of LHC data
  within two-Higgs-doublet models}},  {\em JHEP} {\bf 12} (2013) 095,
  [\href{http://xxx.lanl.gov/abs/1310.7941}{{\tt arXiv:1310.7941}}].

\bibitem{Buchmuller:1985jz}
W.~Buchmuller and D.~Wyler, {\it {Effective Lagrangian Analysis of New
  Interactions and Flavor Conservation}},  {\em Nucl. Phys. B} {\bf 268} (1986)
  621.

\bibitem{AguilarSaavedra:2008zc}
J.~Aguilar-Saavedra, {\it {A Minimal set of top anomalous couplings}},  {\em
  Nucl. Phys. B} {\bf 812} (2009) 181--204,
  [\href{http://xxx.lanl.gov/abs/0811.3842}{{\tt arXiv:0811.3842}}].

\bibitem{Kane:1991bg}
G.~L. Kane, G.~Ladinsky, and C.~Yuan, {\it {Using the Top Quark for Testing
  Standard Model Polarization and CP Predictions}},  {\em Phys. Rev. D} {\bf
  45} (1992) 124--141.

\bibitem{Agashe:2014kda}
{\bf Particle Data Group} Collaboration, K.~Olive et~al., {\it {Review of
  Particle Physics}},  {\em Chin.Phys.} {\bf C38} (2014) 090001.

\bibitem{Aaltonen:2011rj}
{\bf CDF Collaboration} Collaboration, T.~Aaltonen et~al., {\it {Search for a
  Higgs Boson in the Diphoton Final State in $p\bar{p}$ Collisions at $\sqrt{s}
  = 1.96$ TeV}},  {\em Phys. Rev. Lett.} {\bf 108} (2012) 011801,
  [\href{http://xxx.lanl.gov/abs/1109.4427}{{\tt arXiv:1109.4427}}].

\bibitem{Abazov:2011ix}
{\bf D0 Collaboration} Collaboration, V.~Abazov et~al., {\it {Search for the
  standard model and a fermiophobic Higgs boson in diphoton final states}},
  {\em Phys. Rev. Lett.} {\bf 107} (2011) 151801,
  [\href{http://xxx.lanl.gov/abs/1107.4587}{{\tt arXiv:1107.4587}}].

\bibitem{Abbiendi:2013hk}
{\bf LEP, DELPHI, OPAL, ALEPH, L3} Collaboration, G.~Abbiendi et~al., {\it
  {Search for Charged Higgs bosons: Combined Results Using LEP Data}},  {\em
  Eur. Phys. J.} {\bf C73} (2013) 2463,
  [\href{http://xxx.lanl.gov/abs/1301.6065}{{\tt arXiv:1301.6065}}].

\bibitem{Gunion:1989we}
J.~F. Gunion, H.~E. Haber, G.~L. Kane, and S.~Dawson, {\it {THE HIGGS HUNTER'S
  GUIDE}},  {\em Front. Phys.} {\bf 80} (2000) 1--448.

\bibitem{Khachatryan:2015baw}
{\bf CMS} Collaboration, V.~Khachatryan et~al., {\it {Search for a low-mass
  pseudoscalar Higgs boson produced in association with a $b\bar{b}$ pair in pp
  collisions at $\sqrt{s} =$ 8 TeV}},  {\em Phys. Lett.} {\bf B758} (2016)
  296--320, [\href{http://xxx.lanl.gov/abs/1511.03610}{{\tt
  arXiv:1511.03610}}].

\bibitem{Celis:2013rcs}
A.~Celis, V.~Ilisie, and A.~Pich, {\it {LHC constraints on two-Higgs doublet
  models}},  {\em JHEP} {\bf 1307} (2013) 053,
  [\href{http://xxx.lanl.gov/abs/1302.4022}{{\tt arXiv:1302.4022}}].

\bibitem{Haber:1978jt}
H.~Haber, G.~L. Kane, and T.~Sterling, {\it {The Fermion Mass Scale and
  Possible Effects of Higgs Bosons on Experimental Observables}},  {\em
  Nucl.Phys. B} {\bf 161} (1979) 493.

\bibitem{Hall:1981bc}
L.~J. Hall and M.~B. Wise, {\it {Flavor changing Higgs boson couplings}},  {\em
  Nucl.Phys. B} {\bf 187} (1981) 397.

\bibitem{Donoghue:1978cj}
J.~F. Donoghue and L.~F. Li, {\it {Properties of Charged Higgs Bosons}},  {\em
  Phys.Rev. D} {\bf 19} (1979) 945.

\bibitem{Arhrib:2016vts}
A.~Arhrib and A.~Jueid, {\it {$tbW$ Anomalous Couplings in the Two Higgs
  Doublet Model}},  {\em JHEP} {\bf 08} (2016) 082,
  [\href{http://xxx.lanl.gov/abs/1606.05270}{{\tt arXiv:1606.05270}}].

\end{thebibliography}\endgroup

\end{document}